# RISM - Reputation Based Intrusion Detection System for Mobile Adhoc Networks

Animesh Kr Trivedi[1], Rishi Kapoor[1], Rajan Arora[1], Sudip Sanyal[1] and Sugata Sanyal[2]
[1]Indian Institute of Information Technology, Deoghat, Jhalwa, Allahabad (U.P.), India
{aktrivedi_b03, rkapoor_b03, rarora_b03,ssanyal}@iiita.ac.in
[2]School of Technology and Computer Science, Tata Institute of Fundamental Research, India
sanyal@tifr.res.in

*Abstract*—**This paper proposes a combination of an Intrusion Detection System with a routing protocol to strengthen the defense of a Mobile Adhoc Network. Our system is *Socially Inspired*, since we use the new paradigm of Reputation inherited from human behavior. The proposed IDS also has a unique characteristic of being *Semi-distributed*, since it neither distributes its Observation results globally nor keeps them entirely locally; however, managing to communicate this vital information without accretion of the network traffic. This innovative approach also avoids void assumptions and complex calculations for calculating and maintaining trust values used to estimate the *reliability of other nodes' observations*. A robust *Path Manager* and *Monitor* system and *Redemption* and *Fading* concepts are other salient features of this design. The design has shown to outperform normal DSR in terms of Packet Delivery Ratio and Routing Overhead even when up to half of nodes in the network behave as malicious.**

## I. INTRODUCTION

The flexible structure and volatile environment of Mobile Adhoc NETworks (MANETs) causes significant node misbehaviour. It not only degrades the overall Network performance, but also becomes difficult to detect an Intruder, on grounds of mobility of the Nodes. Thus, there is a serious need for a robust *Intrusion Detection System* (IDS) for MANETs.

Some fundamental problems of MANETs must be kept in mind while designing any security solution. *First*, it is often very hard to differentiate intrusions and normal operations or conditions in MANETs because of the dynamically changing topology and volatile physical environment. *Second*, mobile nodes are autonomous units that are capable of roaming independently in an unrestricted geographical topology. This means that nodes with inadequate physical protection can be captured, compromised, or hijacked. *Third*, decision-making in ad-hoc networks is usually decentralized and many ad-hoc network algorithms rely on the cooperative participation of all nodes. Most ad-hoc routing protocols are also cooperative in nature and hence can be easily misguided by false routing information. Without any counter policy, the effects of misbehavior have been shown to dramatically decrease network performance. In this paper, we have proposed a new technique based on Reputation for efficiently solving the problem of intrusion detection.

The next section entails a discussion of some related efforts which is followed by the RISM system design in section III. Section IV describes protocol and the simulation results are given in the section V. The last section presents some concluding remarks.

## II. RELATED WORK

Reputation-based systems are a new paradigm and are being used for enhancing security in different areas. These systems are lightweight, easy to use and are capable of facing a wide variety of attacks, as long as they are observable [1]. Misbehavior Detection and Reputation based intrusion detection Systems may be distributed or local. Here, fully distributed means that information regarding reputation change of a node is immediately propagated in whole network. In the latter case, called Local Reputation based Systems, nodes are fully dependent on their own personal view about reputation and behavior of other nodes.

Distributed IDS protocols either rely only on first-hand information or on positive second-hand information. CONFIDANT proposed by Buchegger and Le Boudec [2] and CORE [3] proposed by P. Michiardi fall into this category. Some basic problems with this approach of global reputation systems are:

- Every node has to maintain O(n) reputation information where n is number of nodes in network.
- Extra traffic generation in reputation exchange.
- Extra computation in accepting indirect reputation information (secondhand information) esp. Bayesian Estimation.
- Security issues in reputation exchange such as reputation data packets can be modified.

The Second type of IDS is one that solely depends upon the first hand observation for reputation maintenance which may be called Local Systems. OCEAN [4] by Bansal and Baker falls in this category. In these systems, nodes make routing decisions based only on the direct observations of their neighbor nodes. This eliminates most of the trust manager complexity, but, in a highly mobile ad-hoc network, it might not be appropriate to depend solely upon personal observation. Also, using second-hand information can significantly accelerate the detection and subsequent isolation of malicious nodes in mobile ad-hoc networks [5].

### III. RISM-OVERVIEW

As previously stated, RISM design is based on the Reputation paradigm and possesses a *Semi-distributed* nature. The term semi-distributed is used for the system observation which is neither restricted locally to our self nor immediately propagated it to the whole network as is the case in true distributed systems like CONFIDANT. The design has been kept very simple keeping in mind the amount of traffic already in the network and the critical amount of battery and computational power individual nodes possess. RISM system runs on every node in network and consists of the following modules:

#### A. Monitor

The Monitor, holds the responsibilty of monitoring activities in the Neighborhood using PACKs (Passive ACKnowledgements) which have been provided as a feature in the DSR protocol specifications [6] as Promiscuous Mode. Every node registers all the *data packets* sent by it to next node and when it receives packets in promiscuous mode, it matches those to the queue of registered packets present in its buffer. After a fixed time interval -termed as the *Timing Window*, nodes make a log of number of packets for which they haven't received acknowledgment in the form of PACK and communicate it to the reputation manager. In existing Reputation systems every packet is kept waiting for its PACK for a fixed time interval, in contrast we have used the concept of Timing Window, which gives us flexibility of checking timeout on fixed intervals (that is after every second) rather checking it on basis of each individual packet's timeout. Monitor maintains a log of activity of next neighbor for each Window and sends it to Reputation manager. The Monitor also takes into consideration the *congestion state* of nodes, which shall be explained in next subsection.

#### B. Reputation System

Reputation system module assigns and maintains reputation of different nodes. Reputation of any node can change due to:
- Self observation
- WARNING Message, issued by neighboring nodes
- Avoid List, appended to the RREQ/RREP header

All the three ways have associated reputation weights with maximum weightage to self observation. The reputation is updated after every Time Window and information is communicated by means of avoid lists, thereby avoiding much of network overhead [5]. A node may be tagged as *Normal*, *Suspicious* or *Malicious* depending on the reputation value associated with it. A node may degrade its reputation by degrading its performance or it may get *Positive Appraisal* on normal behavior. After each timing window Reputation system receives activity log of next hop neighbor from monitor with number of packets for which it does not receive PACK, called as Missing or Dropped Packets. The number of missing packets is then compared with the *MaliciousDropThreshold* and if it is less, then the reputation manager gives positive performance appraisal else negative. Unlike existing systems RISM system does not have a fixed MaliciousDropThreshold, instead we have introduced the concept of *Congestion Parameter*:

$$Congestion Parameter = \frac{Current\ queue\ status}{Total\ queue\ length} \quad (1)$$

with the assumption that next nodes is also in same congestion state as the node in contention, misbehavior drop threshold is dynamically decided after each timing window as

$$MaliciousDropThreshold = MaxPacketRate \times CongestionParameter \quad (2)$$

In order to deal with the attacks on a typical reputation system, like those of 'Collusion of liars' and 'false warning message', the system has a policy that *nodes can be categorized malicious only by Self-observation*. It helps in nullifying the effect of attacks of false warning messages and collusion of liars as the false warning messages they spread, can only decrease reputation of nodes to a certain extent, which is termed as a Suspicious threshold. After which, the system solely depends upon its self observation. Warning messages and Avoid List are only effective above suspicious threshold. Thus, a node is declared malicious eventually through self observation only. Whenever any node has a reputation in that category and the deciding node receives any new warning message or avoid list appearance, the system performs a *Knock Test*, a test designed solely for checking authenticity of a node against whom the deciding node constantly receives such information. In this test, the deciding node generates a fake data packet and forwards it to the node in question. If next nodes forward this packet successfully, then system gives it a performance appraisal and clears its account else the node is declared as malicious. This test can be performed only on nodes in strict or the immediate neighborhood. If the node in question is actually malicious, then it is likely to drop the test packet and hence, monitor shall report its activity to reputation manager and appropriate steps are taken; else, its packets are forwarded and its account cleared.

#### C. Path Manager

The path manager performs trivial path management functions in collaboration with DSR core. Path ranking according to path priority forrmula as given below, updating path-cache on diffrent events like when new nodes declared as malicious or malicious node is taken back in network, taking decision on recieving route request or traffic from malicious nodes are few responsibilities of path manager.

The following function may be used to decide the path priority if need arises:

$$Path\ Priority \propto \frac{1}{|Minimum\ reputation\ of\ Node\ in\ path| \times no\ of\ hops} \quad (3)$$

#### D. Redemption And Fading

*Redemption* and *Fading* are included in design of RISM to allow nodes previously considered malicious to become part of network again as ad-hoc networks run on cooperation and collaboration of peer nodes and no one gets benefited without cooperating with each other. Knock test is very crucial for nodes in suspicious category and node may fail this test due to various reasons like transient link failures, congestion or resetting of the network interface etc. and once they fail this test, they are declared as malicious. To account for these problems, RISM uses the *Fading* mechanism. After a certain inactivity period, the reputation of a node is decreased by a certain predefined fading rate and finally the node is moved from the malicious list to middle of suspicious category. But, the node is not given *Neutral Rating* [7] so that if the node again misbehaves, it is immediately put in malicious list and all transactions through that node are blocked. Here, inactivity period means no appearance in any WARNING messages or avoid list.

## IV. PROTOCOL DESCRIPTION

Initially all the nodes are given Neutral Rating and rating is incremented and decremented on receiving positive and negative events respectively from the Monitor component. Rating of a node never goes above 0 (Neutral rating) as this may lead to favoring of certain nodes in the path, hence causing depletion or draining of resources of these nodes. Thus, preventing attacks of type when a node first earns reputation and then starts dropping packets. A node can declare a neighborhood node malicious only through its self observation. As shown in the figure 2, each node monitors the behavior of its next-hop neighbors through promiscuous mode. All monitoring events are done in a particular timing window after which, reports are submitted to reputation manager for updating the reputation value (transition 2.2 in Fig. 1). When a node crosses the malicious threshold, it is declared as malicious and a WARNING message is broadcasted *only* for immediate neighbors (transition 3.1 in Fig. 1). Thus, protecting the network from reputation flooding. On receiving WARNING message, each node decrements reputation value of reported node with associated weight -as described in reputation manager (transition 3.2 in Fig. 1). Whenever reputation of neighborhood node falls below suspicious threshold and a WARNING message is received, then Knock test is performed on that node. If node passes the test, then it is put in the middle of suspicious category, else it is categorized as malicious and WARNING is broadcasted.

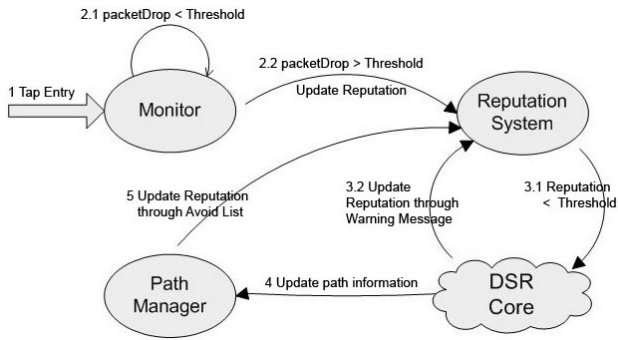

Fig. 1. RISM System Behaviour

To make it possible to avoid routes containing nodes in the malicious list, RISM adds variable-length field to the DSR Route-Request Packet (RREQ) called the Avoid-List [5], a concept taken from OCEAN. The avoid list is a list of nodes that the source of RREQ wants to avoid in its future routes. During RREP, only a path with clean nodes is preferred over those containing *suspicious* nodes and *malicious* nodes. Replies from such nodes are also dropped and nodes do not process request and/or forward data packets from these nodes. Any node receiving an RREQ checks the RREQ avoid list and depending upon the avoid list, nodes having their name in list may be simply dropped. In case if they reply to it by appending them to path, then, at route reply phase, the reply is suppressed if path contains any node listed in avoid list or in the locally-maintained faulty list. On re-broadcasting a RREQ, a node appends its faulty list to the avoid list of the RREQ packet avoiding repetition of already appended nodes. If during traffic flow, a new node is declared malicious, then, all paths containing that node are deleted from route cache and a route error is generated, stating that their link to the destination node has been broken. Neighbors after recieving a route error clear the activity log of the node which generates a route error, from the current timing window.

## V. IMPLEMENTATION/SIMULATION

This section describes the simulation environment setup on the Network Simulator (version 2.29) and compares the throughput of RISM system against a Defenseless DSR protocol. The random waypoint model is selected as a mobility model in a 1000 x 1000 m$^2$ rectangular field with five different pause times: 0, 100, 300, 600 and 900 seconds. There are two type of scenario setups having a total of 10 and 20 nodes, with varying 10 to 100% malicious nodes participating in each. We use max 5 and 10 CBR connections for 10 and 20 nodes respectively, sending 64 bytes packets with a 4 pkts.s$^{-1}$ sending rate. The malicious nodes behave in the following way: dropping an average of 99% of the CBR-connection packets (data packets). However, malicious nodes do not drop DSR routing packets like route request, route reply or error as it is assumed that they always want to be part of the network.

**Table 1: Fixed Parameters**

| Parameter | Level |
|---|---|
| Area | 1000 m x 1000 m |
| Speed | uniformly distributed between 0 and 10 m.s$^{-1}$ |
| Radio Range | 250 m |
| Placement | uniform |
| Movement | random waypoint model |
| MAC | 802.11 |
| Sending capacity | 2 Mbps |
| Application | CBR |
| Packet size | 64 B |
| Simulation time | 900 s |

A malicious node which drops all the packets is less susceptible to harm the network because during the route discovery process, it would not be able to include itself in any of the routes. Thereby refraining itself from affecting throughput of the whole network.

For evaluating the performance of RISM, we consider the following metrics: Packet Delivery Ratio and Routing Overhead. Packet delivery ratio is simply calculated as ratio of data packets received to data packets send. For routing overhead, we consider the ratio of number of control packets generated (request, reply and error) to the number of data packets sent. Routing overhead ratio gives approximate number of control packets for each data packet send. This ratio should not be significantly large as compared to that of normal DSR. Unless otherwise specified, the experiments are repeated ten times with varying random seed. The fixed parameters for the simulation are listed in Table 1. Finally, CBR (Constant Bit Rate) has been chosen for traffic. The scenario and traffic connection files have been randomly generated using CMU's cbrgen and setdest utility.

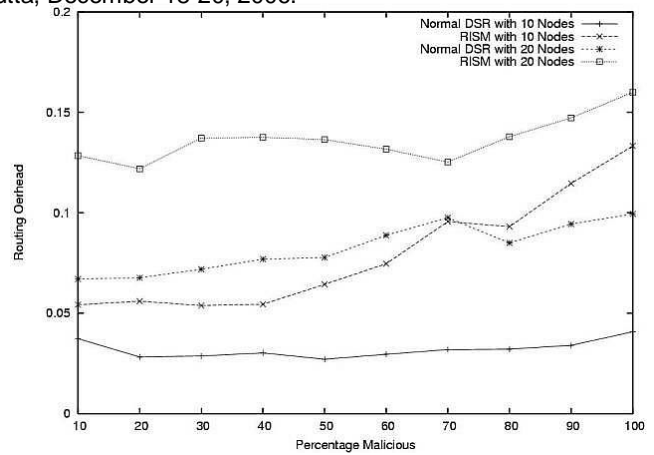

Fig. 3. Routing Overhead Comparison

## VI. RESULT

Results after simulation of Packet Delivery Ratio are shown in Figure (2). Our system RISM performs better than normal

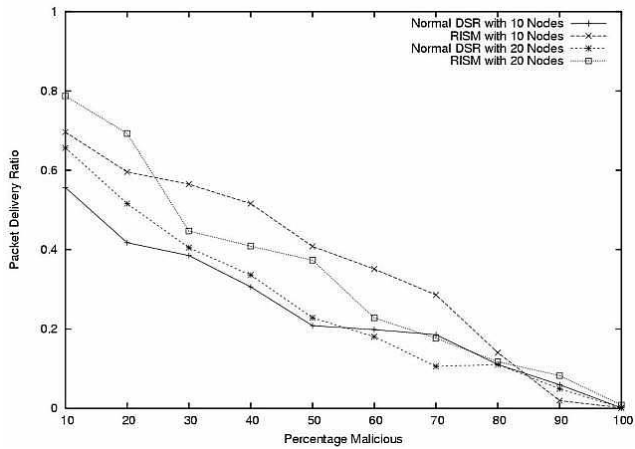

Fig. 2. Packet Delivery Ratio Comparison

DSR taken average over 10 iterations with different pause times. RISM System performance is significantly better than normal DSR when percentage of malicious node less than 40. After that, the system performance starts to deteriorate and significantly falls after 70%. But for networks with 70% or more nodes malicious we can simply discard the network. There is no need to establish trust relationship and links among the nodes when 7 out of 10 are known to be malicious.

Figure (3) shows routing overhead of RISM protocol as compared to normal DSR. Number of Control messages in network are important as more of packets are there, more time is wasted in establishing routes and less the data packets are send. RISM system performs better than normal DSR with not much extra added routing overhead. This extra routing overhead is generated because whenever a new node is declared as malicious, a route error is generated and link is broadcasted as broken. Now some more time is consumed to establish new link. That is crucial to the IDS performance.

## VII. CONCLUSION AND FUTURE PERSPECTIVE

Mobile ad hoc networks have a number of significant security issues which cannot be solved alone by Intrusion detection systems. In this paper we have critically examined the existing systems and outlined their strength and shortcomings. This analysis forms the basis for the proposed socially inspired reputation based Intrusion Detection System, RISM for mobile ad-hoc network. Detailed simulations are carried out for RISM on ns-2 for their adequacy and performance evaluation. Congestion parameter, Knock test and Timing window are some new concepts that are introduced in this paper, which require some further analysis.